\documentstyle[twoside,fleqn,espcrc2,amstex,epsf]{article}

\newcommand\csw{c_{\rm SW}}
\newcommand\psibar{\overline\psi}
\newcommand\overlrarrow[1]{\overset{\leftrightarrow}{#1}}
\newcommand\Dslash{{\smash{\raise 0.16ex \hbox to 0pt{\kern 0.22em $/$
\hss}}D}}

\title{%
\vspace{-3.1cm}
\begin{flushleft}
       {\normalsize DESY 96--142}    \\[-0.2cm]
       {\normalsize HUB--EP--96/23}  \\[-0.2cm]
       {\normalsize July 1996}   \\
\end{flushleft}
       \vspace{0.7cm}
First results with non-perturbative fermion improvement%
\thanks{Talk presented by P.W. Stephenson at Lat96, St. Louis.}}

\author{M.~G\"ockeler%
           \address{Institut f\"ur Theoretische Physik E, RWTH  Aachen,
                    D-52056 Aachen, Germany}$^{\hspace{-0.02cm},%
\hspace{-0.1cm}}$
           \address{H{\"o}chstleistungsrechenzentrum HLRZ,
                    c/o Forschungszentrum J{\"u}lich, D-52425 J{\"u}lich,
                    Germany},
        R.~Horsley%
           \address{Institut f\"ur Physik, Humboldt-Universit\"at zu Berlin,
                    Invalidenstra{\ss}e 110, D-10115 Berlin, Germany},
        M.~Ilgenfritz$^{\rm c}$,
        H.~Perlt%
           \address{Fak.\ f.\ Physik und Geowiss., Universit\"at Leipzig,
                    Augustusplatz 10--11, D-04109 Leipzig, Germany},
        H.~Oelrich$^{\rm b}$,
        P.~Rakow$^{\rm b}$,
        G.~Schierholz$^{\rm b,\hspace{-0.1cm}}$
           \address{Deutsches Elektronen-Synchrotron DESY,
                    Notkestra{\ss}e 85, D-22603 Hamburg, Germany},
        P.~Stephenson%
           \address{DESY-IfH Zeuthen, D-15735 Zeuthen, Germany}
        and
        A.~Schiller$^{\rm d}$}

\begin{document}

\begin{abstract}
We present initial results for light hadron masses and nucleon structure
functions using a recent proposal for eliminating all $O(a)$ effects
from Wilson fermion simulations in the quenched approximation.  With
initially limited statistics, we find a much more linear APE plot and
a value of the axial coupling $g_A$ nearer to the experimental point
than with comparable runs using unimproved Wilson fermions.
\end{abstract}

\maketitle

\section{NON-PERTURBATIVE IMPROVEMENT}

It is now common to seek to improve Wilson fermions by the addition of
a term of the form
\begin{equation}
S_{\rm SW} = \frac{i}{2} \kappa g \csw a
  \sum_{x}\psibar(x)\sigma_{\mu\nu}F_{\mu\nu}(x)\psi(x) \label{eq:sw}
\end{equation}
in order to reduce $O(a)$ lattice cut-off effects present in
the action.  The original proposal~\cite{SW} was to use $\csw=1$,
which corresponds to the tree-level perturbation theory result; this
elminates terms up to $O(ag^2)$.  The elimination of
tadpole contributions requires a value of
$\csw$ some $50\%$ larger depending on the coupling~\cite{Michael}, though
the leading corrections here remain formally $O(ag^2)$.

It has recently been suggested~\cite{Rainer} that $\csw$ can be chosen
via some suitable physical condition so as to remove all $O(a)$
effects.  Here we present initial results using this action as part of
our project for a non-perturbative calculation of nucleon matrix
elements~\cite{qcdsf}.

\section{IMPLEMENTATION}

Our initial runs used a $16^3\times32$ lattice at $\beta=6.0$ with
$\csw=1.769$~\cite{Rainer}.  Our implementation runs on a Quadrics QH2
parallel computer with an $8\times8\times4$ topology.  The improvement
term in equation~(\ref{eq:sw}) appears in the site-diagonal part of the
action; the major overhead in our case is multiplication by this term
during inversion of the fermion mass matrix.  In our gamma matrix
basis,
\begin{xalignat}{2}
  \gamma_i & = i
  \begin{pmatrix}
    0 & \sigma_i \\ -\sigma_i & 0
  \end{pmatrix},
  & \gamma_4 & =
  \begin{pmatrix}
    1 & 0 \\ 0 & -1
  \end{pmatrix}
\end{xalignat}
we can rewrite this term as
% Oh, great.  If this is summer it must be a write-up where LaTeX
% insists on squashing everything over to the right and covering up
% the equation number.  Now what did I do the last ten times?
\begin{multline}
  1 + \kappa c\sigma\cdot F \equiv
  \begin{pmatrix}
    A & B \\ B & A
  \end{pmatrix} \\
  {}\hskip -3em
  \equiv \frac{1}{2}
  \begin{pmatrix}
    1 & -1 \\ 1 & 1
  \end{pmatrix}
  \begin{pmatrix}
    A+B & 0 \\ 0 & A-B
  \end{pmatrix}
  \begin{pmatrix}
    1 & 1 \\ -1 & 1
  \end{pmatrix}\label{eqn:swterm}
\end{multline}
(where $A$, $B$ are $6\times6$ matrices, i.e.\ two-spinors with
colour) so that instead of a $12\times12$ multiplication we have two
$6\times6$ multiplications and two inexpensive co-ordinate
transformations.  This reduces the overhead for the improvement in the
inverter from $45\%$ to $30\%$. Also, the inverse of the term in
equation~(\ref{eqn:swterm}) is required on half the lattice due to the
red-black preconditioning; we now have to invert two $6\times6$
instead of a $12\times12$ matrix.  However, this is only required once
for each propagator inversion, so is a less significant factor.

At $\beta=5.7$ we see large numbers of configurations where the
fermion matrix inversion fails to converge even after 2000 minimal
residual sweeps.  For $\csw>1$ these amount to more than $5\%$ of the
total.  For $\beta<6.0$, no non-perturbatively calculated $\csw$ is
available~\cite{Rainer} for essentially this reason.

\section{HADRON MASSES}

\begin{figure}[hbt]
\vspace*{-1.6cm}
\hspace*{-0.7cm}
\epsfxsize=8.7cm \epsfbox{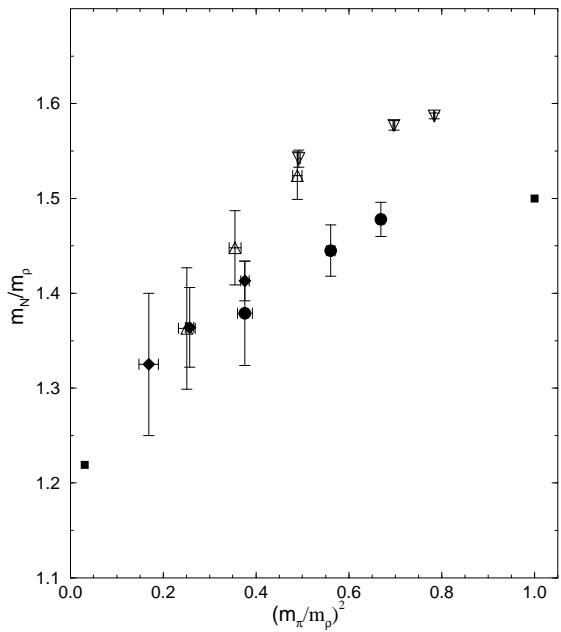}
\vspace*{-1.60cm}
\caption{\footnotesize The APE plot for the unimproved (unfilled
  symbols) and 
  $\csw=1.769$ improved (filled) data.  The lattices used are
  $16^3\times32$  and $24^3\times32$ 
  at $\beta=6.0$.  The physical and heavy quark points are shown as squares.}
\vspace*{-0.8cm}
\label{apeplot}
\end{figure}

We have calculated pion, rho and nucleon masses for $\kappa=0.1324$,
$0.1333$ and $0.1342$ from 125 configurations, and on a $24^3\times32$
lattice at $0.1342$, $0.1346$ and $0.1348$ with similar statistics.
The APE plot is shown in figure~\ref{apeplot}, where we compare the
results with previous runs using $\csw=0$ at $\kappa=0.1515$, $0.1530$
and $0.1550$ with large statistics (some 5000 data points) on the
smaller lattice and $0.1550$, $0.1558$ and $0.1563$ again
with around 100 configurations on the larger lattice (the high
statistics and light quark results are both previously unpublished).
The $\kappa$'s have been chosen to lead to similar pion masses in the
improved and unimproved data sets; we also find that the nucleon
masses are similar.  However, the rho mass with improved fermions is
some $20\%$ larger for the higher mass points, which are thus shifted
towards the origin in comparison with the Wilson case.

As may be seen, we find that the unimproved Wilson hadron masses
eventually converge with the improved values, and indeed the
extrapolated values are comparable, although the extrapolation of the
ratios deviates considerably more from the linear in the former case.  We
have also found extrapolation of the individual masses to the chiral
limit to be more linear for the improved fermions.

\section{STRUCTURE FUNCTIONS}

For a full non-perturbative calculation of the structure functions we
need two additional ingredients.

First, we require operator improvement which needs to be calculated
separately for each operator.  The perturbative case has
\begin{gather}
\begin{split}
  &V \equiv \psibar\gamma_\mu\psi \to \psibar\gamma_\mu\psi \\
  &\qquad+ b_Va(-\psibar\overlrarrow{D}_\mu\psi
  -\partial_\nu\psibar\sigma_{\mu\nu}\psi),
\end{split} \\
\begin{split}
  &A \equiv \psibar\gamma_\mu\gamma_5\psi \to
  \psibar\gamma_\mu\gamma_5\psi \\
  &\qquad+ b_Aa(\partial_\mu\psibar\gamma_5\psi 
  + \psibar\sigma_{\mu\nu}\overlrarrow{D}_\nu\gamma_5\psi);
\end{split}
\end{gather}
in this preliminary calculation we have set $b_V = b_A = 1/2$,
corresponding to the `rotation' valid in the $\csw=1$ case.
Non-perturbatively, we need to determine the coefficients of each term
separately.

Secondly, we also need to determine the renormalisation constants
$Z_O$ non-perturbatively.  At present we use a perturbative
calculation assuming the $b_V=b_A=1/2$ rotation but with
$\csw=1.769$.  A point to note is that definitions of $Z_O$ vary,
depending on whether the residual effect of the rotation in the chiral
limit is absorbed into $Z_O$ or not: we follow Martinelli et
al.~\cite{Martinelli}.  This is different to the prescription being
used by the ALPHA collaboration for these calculations, hence their
results presented at this conference~\cite{Hartmut} are not
immediately applicable to us.  The calculation of the $Z_O$ is then
similar to that presented in reference~\cite{Borrelli}.

A fully non-perturbative calculation of the renormalisation constants is
currently in progress.

\subsection{Local vector current}

We have calculated the local vector current and extrapolated to the
chiral limit, where the perturbative value for $Z_V$ was calculated.
As this current is conserved in the continuum we require only to
recover the values 2 for the up quark and 1 for the down quark content
of the proton.  This acts as a test of the consistency of our
procedure.  The results are $2.0(3)$ and $1.0(2)$, which are entirely
satisfactory.

\subsection{Axial vector current}

\begin{figure}[thb]
\vspace*{-2.4cm}
%\hspace*{-1.50cm}
\epsfxsize=10.0cm \epsfbox{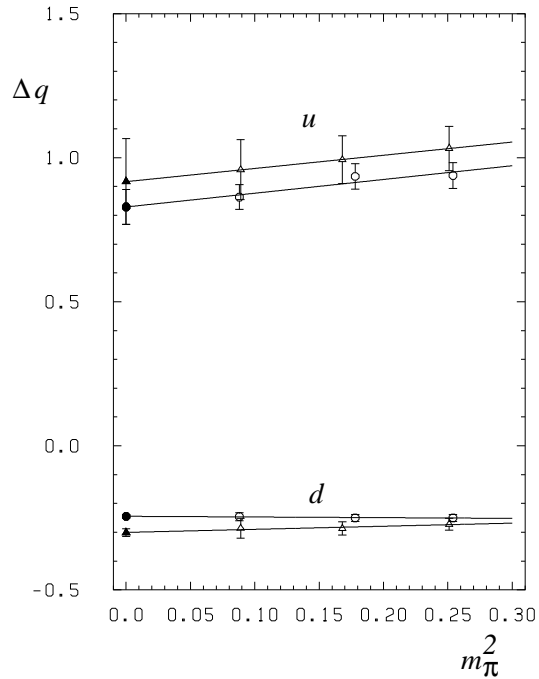}
\vspace*{-3.25cm}
\caption{\footnotesize The axial vector currents for up and down
  quarks, $\Delta u$ and $\Delta d$, for both improved ($\triangle$)
  and unimproved ($\circ$) fermions, including the factor $Z_A$
  calculated in the chiral limit (solid points).}
\vspace*{-0.75cm}
\label{axial}
\end{figure}

The axial vector current $\Delta q$ is shown in figure~\ref{axial}.
This was formerly (in the naive quark model) connected with the spin
contribution of the quarks, though emphasis is now on the violation of
the OZI rule implied by the low values for the flavour singlet
operator~\cite{Dis}.

Our main interest is in the combination $g_A=\Delta u-\Delta d$ (the
axial coupling) where disconnected fermion loops which we have not
calculated cancel out.  The experimental value is 1.26.  With
the improvement, our calculated quantity has changed from $1.07(9)$ to
$1.22(14)$.  This is better but suffers from large errors.

\section{CONCLUSIONS}

% Improvement works, everything is wonderful, it's Christmas Eve and oh,
% look, it's started to snow and Santa Claus is coming.

We have performed a first QCD calculation with Wilson fermions
improved up to $O(a^2)$.  Hadron masses with
improved fermions at $\beta=6.0$ extrapolate more linearly to the
chiral limit than with ordinary Wilson fermions.

Our preliminary analysis of the local vector current in the nucleon
shows the procedure to be consistent, and our results for the axial
vector indicate a promising trend in our value for the axial coupling
despite low statistics.  A fully non-perturbative analysis remains to
be done.

The work was done on the APE Quadrics computers at DESY-IfH, Zeuthen,
Germany.  We thank the computer center for technical help.

\end{document}